# Laser-based underwater transfer of radio-frequency clock signal with electronic phase compensation

DONG HOU,[1,*] GUANGKUN GUO,[1] DANIAN ZHANG,[1] AND FUYU SUN[2]

[1]*Time & Frequency Research Center, School of Automation Engineering, University of Electronic Science and Technology of China, Chengdu, Sichuan 611731, China*
[2]*National Time Service Center, Chinese Academy of Science, Xi'an, Shanxi 710600, China*
*Corresponding author: houdong@uestc.edu.cn

We demonstrate a laser-based underwater transfer of radio frequency clock signal with a phase compensation technique. With this frequency transfer scheme, a 100 MHz signal has been transferred over 5 m underwater link. Timing jitter power spectral density, timing fluctuations and instabilities were all measured to evaluate the quality of the transferred frequency signal. The experimental results show the root-mean-square (RMS) timing fluctuations of the transferred frequency signal over 5 m underwater link with and without phase compensation are 2.1 ps and 9.6 ps within 5000 s, respectively. The relative fractional frequency instabilities are on the order of $5 \times 10^{-13}$ at 1 s and $7 \times 10^{-16}$ at 1000 s for phase-compensated link, and $2 \times 10^{-12}$ at 1 s and $7 \times 10^{-15}$ at 1000 s for uncompensated link, respectively. The proposed frequency transfer scheme has a potential application that disseminating Cs and H-master clocks signal over the underwater transmission link since the stability of the transmission link is superior to these atomic clocks. © 2018 Optical Society of America

*OCIS codes: (060.2605) Free-space optical communication; (010.4455) Oceanic propagation; (120.0120) Instrumentation, measurement, and metrology; (270.2500) Fluctuations, relaxations, and noise.*

http://dx.doi.org/10.1364/PRxxxxxx

## 1. INTRODUCTION

In the last decades, the underwater wireless communication has attracted a remarkable attention as it is capable of providing a higher flexibility and cost-effectiveness than submarine optical fiber communication which can only be used in the data transmission between the static objects in the water [1-3]. Up to now, the underwater wireless communication has been widely used in the data transmission between various underwater vehicles, sensors and observatories [3, 4]. Generally, the traditional wireless underwater communication includes, underwater acoustic and microwave methods [5-8]. Although the current available acoustic scheme which uses acoustic waves for underwater communication can reach a long distances (ranging larger than few hundred meters), it is still limited by the low bandwidth, high transmission losses, time varying multipath propagation, and high latency [5]. The transmission of microwave have a high-bandwidth and high propagation speed. However, the microwave signal is highly attenuated by the water, which significantly limits the distance of the underwater communication (ranging only few meters) [8]. To solve the problem that high-bandwidth and long-distance are both required in the underwater wireless communication, another method of transferring optical signal over underwater link was proposed [3, 9].

In this underwater optical wireless communication, the optical signal is able to be transferred as far as tens or hundred meters. In the past years, the interest towards optical wireless transmission over underwater link has increased as it is capable of providing high data rates with low power over long-distance underwater link. Some prior works show that the high-speed wireless underwater transmission links with blue and green visible lasers have been achieved [10-17]. As the development of the wireless underwater communication, distribution of timing and frequency signal between the underwater objects becomes more and more important, which has many potential applications in underwater metrology [3, 18, 19], for example, submersible synchronization, underwater navigation, underwater sensing, and etc. The timing and frequency signals has been able to be transferred over atmospheric link. Many frequency dissemination experiments show that atmospheric transfers of optical and radio frequency signals with ultra-low timing deviation have been achieved, where phase compensation schemes were proposed to suppress the turbulence-affected timing fluctuation over free-space optical link [20-24]. However, to the best of our knowledge, the transfer of timing and frequency signal over underwater link has not be reported yet.

In this paper, a laser-based underwater transfer of a radio-frequency signal with a phase compensation scheme is demonstrated. In our underwater frequency transfer experiment, we disseminated a 100 MHz signal over a 5 m underwater link. The experimental results show that the RMS timing fluctuations of the transferred frequency signals with and without phase compensation are 2.1 ps and 9.6 ps in 5000 s, respectively; and the relative fractional frequency instabilities of the transmission links with and without phase compensation are on the order of $5 \times 10^{-13}$ at 1 s and $7 \times 10^{-16}$ at 1000 s, and the order of $2 \times 10^{-12}$ at 1 s and order of $7 \times 10^{-15}$ at 1000 s respectively.

## 2. SCHEMATIC OF UNDERWATER FREQUENCY TRANSFER WITH PHASE COMENSATION

In order to transfer a radio-frequency signal from a transmitter to a receiver via an optical carrier over underwater link, the most convenient scheme includes three steps [3], which are directly loading the radio-frequency signal onto the optical carrier, then transferring the light to remote receiver via an direct underwater optical link, and lastly recovering this radio-frequency signal by a photodetector. This process is almost the same as the underwater optical wireless communication, where the optical light with encoded data is directly transmitted from transmitter to receiver via a visible underwater link. In this case, the bit error rate of the decoded data is determined by the signal-to-noise ratio (SNR) of the transmitted signal, and just slightly influenced by the timing fluctuation or drift. However, in a frequency transfer system, the stability of the transmitted radio-frequency signal is directly determined by the timing fluctuation and drift of the optical signal.

The three steps of the underwater transmission of optical signal presented above could introduce excess timing fluctuation and drift into the signal, resulting in the stability degradation of the original radio-frequency signal. However, the strong timing fluctuation and drift introduced by the water turbulence and temperature drift is much higher than that introduced by laser source and the photo-detection electronics. Therefore, for an underwater transfer of a radio-frequency signal in the water environment, the timing fluctuations and frequency instability of the transmission link is dominantly caused by the water turbulence and temperature drift rather than on the transmitter and receiver [25-27]. Literatures reveal that the water turbulence induces variation of the refraction index [28], which could directly lead to excess phase noise or timing fluctuation in the transmitted frequency signal. Here, the spectrum density of underwater optical turbulence can be expressed as $\Phi_n^k(\kappa) = K_3 \kappa^{-11/3}$[28], where $K_3 = X\varepsilon^{-1/3}$ ($X$ expresses the strength of temperature gradient and $\varepsilon$ is the kinetic energy dissipation rate) is the constant that determines the turbulence strength. For underwater conditions, the value of $K_3$ ranges from $10^{-14}$ to $10^{-8}$ m$^{-2/3}$. Note that, $\kappa$ is the scalar spatial frequency (in rad/m). Therefore, the longer the transmission distance is, the higher the excess timing fluctuation of transmitted frequency signal is introduced. In order to improve the quality of the frequency transfer significantly, the timing fluctuations should be suppressed by a phase compensation technique. In this paper, we propose a scheme of laser-based underwater frequency transfer with a low timing fluctuation using a phase compensation technique.

Figure 1 shows the schematic of our laser-based underwater radio-frequency transfer with an electronic phase compensation technique. The optical carrier is provided by a continuous wave (CW) 520 nm diode laser. A highly-stable frequency source, is phase-shifted by a voltage-controlled phase shifter first, and then directly loaded onto the optical carrier via a current amplitude-modulation (AM) scheme. The modulated laser beam is launched into the water from the transmitter. On the receiver, half of the beam is reflected by a half-mirror, and the remaining beam is converted to a radio-frequency signal via a high-speed photodiode for users. The returning beam which transmits the same optical path along the forward underwater transmission link, is detected by another high-speed photodiode on the transmitter. The detected microwave signal from the photodiode is phase-shifted by an identical electronic phase shifter, and then compared with the reference source signal on a phase detector, to generate an error signal which includes the information of timing fluctuation affected by water turbulence. This error signal is fed-back to the two identical phase shifters via a (proportional-integral) PI servo controller, to adjust the phase delay on each phase shifter. When the servo loop is closed, the timing fluctuations affected by the water turbulence can be corrected by the phase compensation technique. Note that, the phase shifters produce the same phase delays, because they are controlled by the same error signal simultaneously. The mechanism of the phase compensation will be explained in detail below.

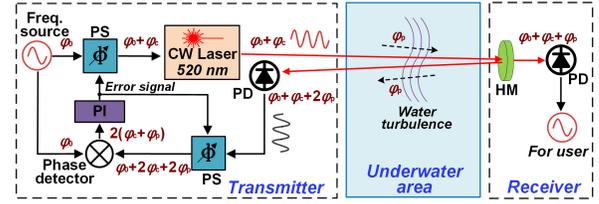

**Fig. 1.** Schematic of the underwater frequency transfer with an electronic phase compensation technique. PS: phase shifter, PD: photodiode, HM: half-mirror, PI: Proportion-Integration controller.

As shown in Fig. 1, we assume that the frequency source generates a radio-frequency signal with an initial phase $\varphi_0$. On the transmitter, this signal is phase-shifted with $\varphi_c$ via a phase shifter first, and then delivered to the receiver over an underwater link. Assuming the water turbulence introduces a phase fluctuation $\varphi_p$ to the transmitted signal over the one-trip transmission link, the total phase delay of the recovered microwave signal on the receiver is given by $\varphi_{total}=\varphi_0+\varphi_c+\varphi_p$. With a half-mirror, half of beam is reflected to the transmitter along the almost same optical path, which will introduces a same turbulence-affected phase fluctuation $\varphi_p$. In this case, the returning microwave signal detected by the photodiode has a phase delay $\varphi_0+\varphi_c+2\varphi_p$ due to the twice turbulence effect. After passing through an identical phase-shifter, another phase delay $\varphi_c$ is introduced to the returned microwave signal. Therefore, the phase delay of the returned microwave signal is given by $\varphi_{returned}=\varphi_0+2\varphi_c+2\varphi_p$. After eliminating the initial phase $\varphi_0$ by comparing it with the reference signal, we get an intermediate frequency signal with the phase error information $2(\varphi_c+\varphi_p)$. Here, this error signal is used to control the two identical phase shifters via a PI servo controller, for compensating the turbulence-affected phase $\varphi_p$. When the servo loop is closed, the phase error $2(\varphi_c+\varphi_p)$ equals zero, and consequently, the timing fluctuation affected by water turbulence for the recovered radio-frequency signal on receiver will be corrected as $\varphi_c = -\varphi_p$. With this electronic phase compensation technique, the timing fluctuation performance and stability of our underwater frequency transfer will be improved significantly, compared to the direct transmission link without phase compensation

## 3. EXPERIMENTAL SETUP

Figure 2 shows the experimental setup of our underwater frequency transfer system using a visible diode laser with the electronic phase compensation. This experiment setup was built in a windows-opened laboratory room. Before the experiment was conducted, we built a water tank with a length of 3.1 m, a width of 0.35 m, and a height of 0.35 m. The water tank was fixed on a sturdy and flat desk, and then we filled it with fresh water. On the transmitter, the optical carrier is provided by a commercial 520 nm green diode laser with an output power of 30 mW, which has a much lower attenuation than red and infrared lasers in water. A 100 MHz microwave signal with a power of 20 mW, generated from a stable signal source (Agilent, E4421B), is directly loaded onto the laser with an amplitude modulation (AM) method. The laser beam is coupled to a collimator, and then launched into the water tank directly. At 2.5 m far away remote site, the beam was sent back to the receiver by a silver-coated two-inch mirror (M1), which forms a total 5 m one-way transmission link in water.

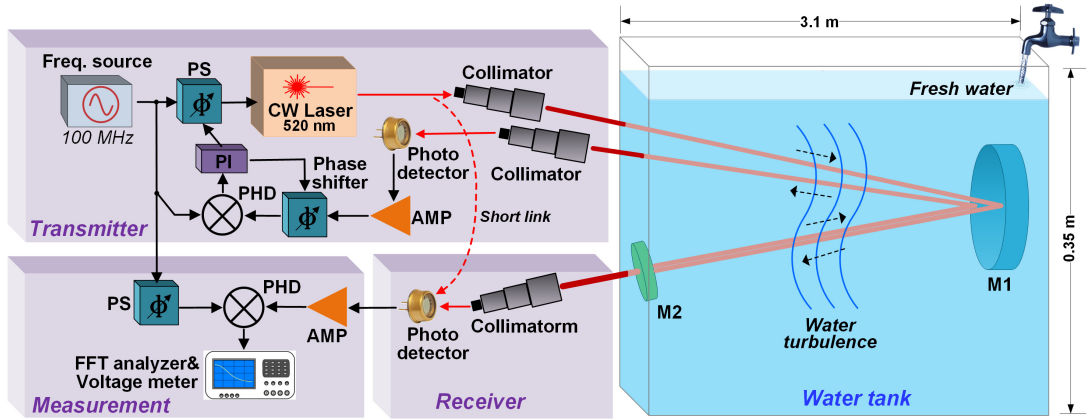

**Fig. 2.** Experimental setup of the underwater frequency transfer with an electronic phase compensation technique. PS: phase shifter, CW: Continuous-wave, PHD: phase detector, PI: proportion-Integration controller, AMP: amplifier.

On the receiver, half of the transmitted beam is reflected to transmitter along the same optical path by a half-mirror (M2). The reflected beam with power of about 1 mW is collected by another collimator and tightly focused onto a high-speed Si photodiode on the transmitter. The converted 100 MHz radio-frequency signal with twice timing fluctuations was amplified, phase shifted, and then mixed with the RF reference source signal to generate the phase error signal. This error signal is then fed-backed to the two identical phase shifters (Mini-circuit, JSPHS-150) via a fast PI servo controller, to compensate the timing fluctuations affected by the water turbulence. On the receiver, with another high-speed Si photodiode, the remaining half of beam with power of about 6 mW is also converted to a radio-frequency signal. This recovered signal, is also amplified, filtered and mixed with the frequency reference source, to produce a DC output. After low pass filtering, the DC signal is delivered to a digital voltage meter (Keysight, 34461A) and a FFT analyzer (HP, HP35665), for estimating the quality of the frequency transmission. Note that, our experimental setup is in an open-window laboratorial environment, and the water tank is not covered. Therefore, in this setup, the water turbulence and temperature drift has an obvious effect on the phase of the transmission signal. In addition, we collected the optical light on the photodetectors as much as possible to acquire the highest signal-to-noise ratio (SNR).

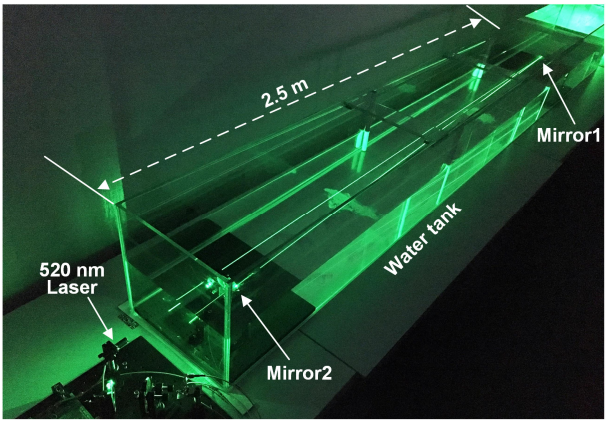

**Fig. 3.** Real photo of experiment setup for underwater frequency transfer.

## 4. EXPERIMENTAL RESULTS AND DISSCUSSION

The frequency transfer experiment was conducted in a normal day. In the experiment, the two collimator were placed as close as possible on the transmitter to obtain an almost identical turbulence effect. The real photo of the underwater frequency transfer experiment is shown in Fig.3. Since the phase compensation technique can suppress the additional timing fluctuations caused by water turbulence, we believe the quality of the transferred 100 MHz frequency signal transfer should be improved significantly compared to the test without phase compensation. Here, we measured the timing jitter power spectral density (PSD) and timing fluctuation of the transferred 100 MHz radio-frequency signals with and without phase compensation.

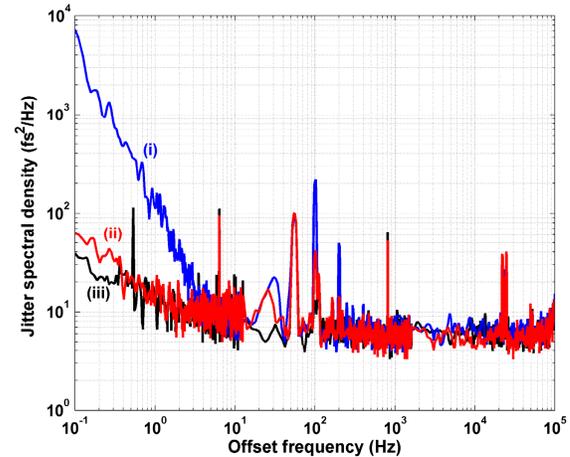

**Fig. 4.** Timing jitter (PSD) results for underwater frequency transfer. Curve (i): Without phase compensation. Curve (ii): With phase compensation. Curve (iii): Result for short link as a measurement floor

The timing jitter PSD results are shown in Fig. 4. Curve (i) and (ii) are the timing jitter PSD data of excess timing jitter of the transferred 100 MHz microwave signal without and with phase compensation, respectively. Curve (iii) shows the background noise floor of the short link (shown Fig. 2), where the modulated light was directly delivered from transmitter to receiver, and not transmitted in water. From Curve (i) and (ii), we can see the timing jitter PSD is suppressed below 5 Hz offset frequency. This indicates the most of short-term timing fluctuation affected by water turbulence is below 5 Hz offset frequency, and it can be suppressed with our phase compensation technique.

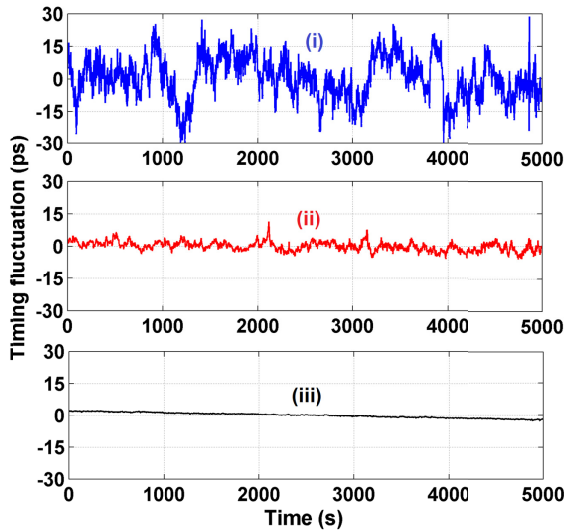

**Fig. 5.** Timing fluctuation results for underwater frequency transfer. Curve (i): Without phase compensation. Curve (ii): With phase compensation. Curve (iii): The result for a short link as a measurement floor. Sample rate is 1 point/second for all curves

The timing fluctuation results are shown in Fig. 5, and each curve was recorded for 5000 seconds. Curve (i) shows the timing fluctuation of the 100 MHz radio-frequency signal recovered from the direct link without phase compensation, and its RMS timing drift is approximately 9.6 ps within 5000 s. Curve (ii) shows the timing fluctuation of the 100 MHz radio-frequency signal recovered from the timing suppressed link with phase compensation, and its RMS timing drift is reduced to about 2.1 ps within 5000 s. Here, we also measured the timing fluctuation of the frequency transfer with a short link as the measurement floor (shown in Fig. 2), where the optical light was not transmitted in the underwater link. Therefore, this timing fluctuation is just attributed by the laser noise and electronic noise of our photonic system. Curve (iii) shows the timing fluctuation of the short link, and its RMS timing drift is approximately 1.3 ps within 5000 s. Based on the comparison between the transmission links with and without phase compensation, we believe that the proposed phase compensation technique effectively suppressed the long-term timing fluctuation.

The instability of the transmission link is very important for a time-frequency distribution system. Therefore, we also calculated the Allan Deviations for the underwater frequency distribution link. Figure 6 shows the instability results for the transferred radio-frequency signal. Curve (i) is the relative Allan Deviation result for the transferred signal without phase compensation, which is calculated out from the sampled data in Fig. 5. From Curve (i), it shows the 5 meters underwater frequency transmission link without phase compensation has a instability of $2 \times 10^{-12}$ for 1 s and $7 \times 10^{-15}$ for 1000 s. Curve (ii) is the relative Allan Deviation result for the transferred signal with phase compensation, and it shows the 5 meters underwater frequency transmission link with phase compensation has a instability of $5 \times 10^{-13}$ for 1 s and $7 \times 10^{-16}$ for 1000 s. With comparison of Curve (i) and (ii), we find that the long-term instability of the underwater transferred frequency signal with phase compensation is reduced for about half order of magnitude at 1 s and one order of magnitude at 1000 s due to the timing suppression. Curve (iii) shows the measurement floor, which was obtained via a short link without any underwater transmission (see Fig. 2). Note that, this curve is merely the lower bound of the instability measurement, and it is just limited by the stability of the frequency source and the electronic noise. This is because the most of water turbulence effect was cancelled. The instability achieved by this proposed underwater frequency transfer system with phase compensation may be quite adequate with some short-range underwater time-frequency distribution applications. For example, with comparison with instability of our transfer result and a commercial Cs clock (5071A) [29] or H-master clock (MHM 2010, standard version) [30], we can find that the instability of our transmission link is lower than the Cs and H-master clocks. This indicates that disseminating a Cs or H clock signals in water environment by using the proposed underwater frequency transfer scheme in this paper is feasible.

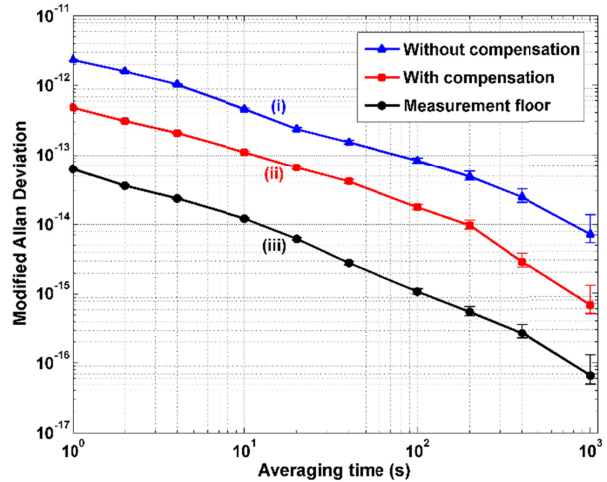

**Fig. 6.** Instability results for underwater frequency transfer, (i) Relative Allan deviation between transferred radio-frequency signal and reference signal without phase compensation; (ii) Relative Allan deviation with phase compensation; (iii) Allan deviation for a short link as measurement floor.

## 5. CONCLUSIONS

We have demonstrated a laser-based radio-frequency transfer over an underwater link with a phase compensation technique. The RMS timing fluctuation for a 100 MHz frequency transfer over a 5-m long underwater link with phase compensation was measured to be approximately 2.1 ps within 5000 s with a fractional frequency instability on the order of $5 \times 10^{-13}$ at 1 s and of $7 \times 10^{-17}$ at 1000 s. The instability of the underwater transmission link with the distance of a few meters is superior to atomic clocks, for example, Cs or H-master clocks, which implies the atomic clock is able to be directly distributed over underwater link in a short-range. Although the distance of the underwater transmission link in our experiment is few meters, this is just limited by the power of the laser and the physical size of the water tank. In the future, a lower timing fluctuation and longer distance underwater frequency transfer will be achieved by loading a microwave signal with higher frequency and using high power laser source.

**Funding Information.** National Natural Science Foundation of China (NSFC) (61871084, 61601084);